\documentclass[conference]{IEEEtran}

\ifCLASSINFOpdf

\else

\fi

\usepackage{hyperref}

\usepackage[dvipsnames]{xcolor}
\newcommand{\myWhite}[1]{\textcolor{white}{#1}}
\usepackage{graphicx}
\usepackage{svg}
\graphicspath{ {./images/} }
\usepackage{ wasysym }
\usepackage{enumitem}
\newlist{RQ}{enumerate}{1}
\setlist[RQ]{label=\textbf{RQ\,\arabic*:},ref={RQ\,\arabic*},itemsep=0ex,topsep=0em,leftmargin=3em}
\usepackage{color,soul}
\soulregister{\myWhite}{1}
\newcommand{\hlc}[2][yellow]{ {\sethlcolor{#1} \hl{#2}} }
\definecolor{myGray}{gray}{0.87}

\usepackage{tcolorbox}
\tcbuselibrary{skins, breakable}
\tcbset{
  width=0.5 \textwidth,
  halign=justify,
  center,
  breakable,
  colbacktitle=white,
  coltitle=black,
  fonttitle=\sffamily\small,%\bfseries
  colback=white,
  sharp corners, % arc=0mm,
  size=title,
  boxrule=0.3mm,
  fontupper=\small,%\ttfamily
}

\newcommand{\rqone}{How do interactions with a GenAI chatbot influence learners’ orienteering behaviors? } % Orienteering in Information Retrieval?
\newcommand{\rqtwo}{How do learners identify information scent in GenAI chatbot responses? } % Information Scent in Information Retrieval?

\begin{document}
%\title{Searching Information Space with Generative AI for Interdisciplinary Learning} 
%\title{Conversations over Clicks: How Chatbots Affect Information Search in Interdisciplinary Learning
%}
\title{\large\textbf Conversations over Clicks: Impact of Chatbots on Information Search in Interdisciplinary Learning
}

\author{\IEEEauthorblockN{Hannah Kim}
\IEEEauthorblockA{\textit{Dept. of Biology}\\
\textit{Temple University}\\
\textit{Philadelphia, PA, USA}\\
hannah.kim0007@temple.edu}
\and
\IEEEauthorblockN{Sergei L. Kosakovsky Pond}
\IEEEauthorblockA{\textit{Dept. of Biology}\\
\textit{Temple University}\\
\textit{Philadelphia, PA, USA}\\
spond@temple.edu}
\and
\IEEEauthorblockN{Stephen MacNeil}
\IEEEauthorblockA{\textit{Dept. of Computer and Information Sciences}\\
\textit{Temple University}\\
\textit{Philadelphia, PA, USA}\\
stephen.macneil@temple.edu}
}

\maketitle

\begin{center}
    \rule{8.4cm}{0.1pt}\\[1ex]
    \fontsize{7.5pt}{9.5pt}\selectfont
    \copyright 2025 IEEE. Personal use of this material is permitted. Permission from IEEE must be obtained for all other uses, in any current or future media, including reprinting/republishing this material for advertising or promotional purposes, creating new collective works, for resale or redistribution to servers or lists, or reuse of any copyrighted component of this work in other works.\\ doi:10.1109/FIE63693.2025.11328556\\
    \rule{8.4cm}{0.1pt}\\[1ex]
\end{center}

\begin{abstract}
This full research paper investigates the impact of generative AI (GenAI) on the learner experience, with a focus on how learners engage with and utilize the information it provides. In e-learning environments, learners often need to navigate a complex information space on their own. This challenge is further compounded in interdisciplinary fields like bioinformatics, due to the varied prior knowledge and backgrounds. In this paper, we studied how GenAI influences information search in bioinformatics research: (1) \rqone; and (2) \rqtwo We adopted an autoethnographic approach to investigate these questions. GenAI was found to support orienteering once a learning plan was established, but it was counterproductive prior to that. Moreover, traditionally value-rich information sources such as bullet points and related terms proved less effective when applied to GenAI responses. Information scents were primarily recognized through the presence or absence of prior knowledge of the domain. These findings suggest that GenAI should be adopted into e-learning environments with caution, particularly in interdisciplinary learning contexts.
\end{abstract} % HK: 184 words
\IEEEpeerreviewmaketitle
%%%%%%%%%%%%

\section{Introduction}

\begin{figure}[h]
    \centering
    \includegraphics[width=\linewidth]{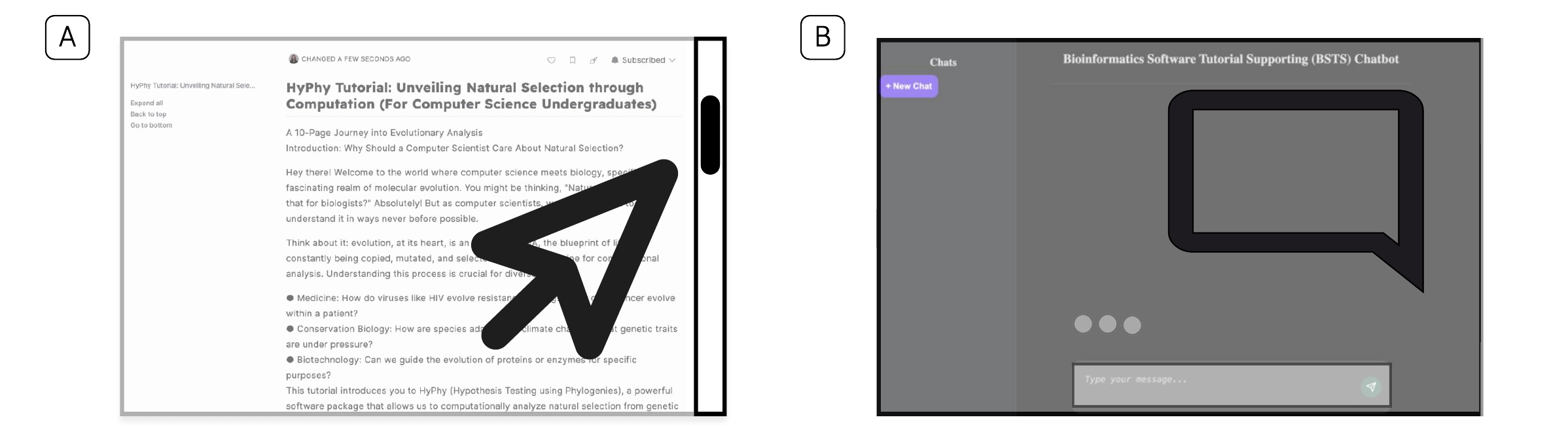}
    \caption{Comparison between a traditional e-tutorial (A) and a GenAI-based conversational e-tutorial (B)}
    \label{fig:1}
\end{figure}

Students are increasingly turning to AI tools like ChatGPT for support  when they encounter challenges \cite{hou2024effects, prather2023robots}. When used with appropriate guardrails, these tools provide many learning benefits \cite{prather2025beyond}. As students continue to shift away from traditional help-resources \cite{hou2025evolving}, it is important to understand how GenAI impacts their ability to find relevant information. In this paper, we investigate how learners’ approaches to information seeking change when moving from traditional e-tutorials to GenAI chatbots, particularly in interdisciplinary contexts.

Prior studies have compared GenAI chatbots to traditional information retrieval systems. While both traditional web search and GenAI chatbots share a query-box interface, the conversational nature of GenAI changes how users formulate queries \cite{wazzan2024comparing}. Traditional retrieval-based chatbots resemble GenAI chatbots in being conversational and ever-present; however, GenAI's capability to generate novel, human-like text with high performance without relying solely on given inputs makes it more versatile \cite{pandey2023comparative}. The potential educational value of conversation-simulating agents has been recognized since the late 1960s, but it is GenAI's performance, versatility, and affordability that have enabled its large-scale adoption in education today \cite{ilieva2023effects}. Accordingly, a common shift in educational technology learners are currently experiencing is the transition from traditional e-tutorials to GenAI chatbots (See Figure \ref{fig:1}).

What makes the dramatic transition from traditional e-tutorials to GenAI logistically practical is the resemblance of GenAI responses--probabilistically generated summaries--to traditional e-tutorials; after all, a tutorial is a summary of knowledge \cite{shulman1987knowledge,walsh2013} accompanied by guiding structures \cite{hirst1971teaching}. However, several key differences between these formats are often overlooked. The dominant user behavior with traditional e-tutorials (Fig. \ref{fig:1}A) is scrolling, whereas interaction with GenAI chatbots (Fig. \ref{fig:1}B) centers on chatting. This difference changes how content is presented to learners: traditional e-tutorials follow a pre-defined order designed by a teacher, while GenAI delivers content dynamically in response to learner queries and model conditions, often at the expense of the teacher-intended instructional structure. Furthermore, GenAI chatbots provide instant, human-like feedback as a core feature unlike traditional e-tutorials. This conversational responsiveness can lend undue credibility to GenAI content. Although some of the challenges posed by these differences may be addressed through curricular integration of AI literacy \cite{long2020ai, knoth2024ai}, they remain underexplored. These concerns are especially relevant in interdisciplinary learning contexts, where the perceived benefits of GenAI, such as personalization and support for self-directed learning \cite{ilieva2023effects}, are particularly compelling.

We investigate the impact of GenAI on e-learning in an interdisciplinary context, using information seeking as the theoretical framework. Information seeking is a purposive human information behavior centered on the acquisition of information \cite{afzal2012management}, making it well-suited for understanding human–AI interactions, since each exchange is initiated by a user query reflecting an underlying purpose. Concepts such as orienteering and information scent build upon this foundation. Orienteering is an information search theory inspired by a sport of the same name, in which participants navigate from a shared starting point to a common destination via multiple potential intermediate locations \cite{o1993orienteering}. In the context of information seeking, orienteering theory suggests that users incrementally transition between diverse search modes (e.g., monitoring, following a plan, and exploring a topic), adapting their approach based on the results encountered until the task is complete \cite{o1993orienteering}. Information foraging theory (IFT) can be used to examine each transition in detail. It models how users allocate limited attention, drawing on evolutionary-ecological models of food foraging \cite{pirolli1999information}. In IFT, a patch of information perceived as valuable is said to have information scent. Because this value is inferred from contextual cues, information scent is inherently proximal and often imperfect \cite{pirolli1999information,budiu2020information}. The information diet refers to the set of information sources perceived as most profitable and prevalent in value based on their information scent \cite{pirolli1999information}. To explore how GenAI chatbots shape knowledge acquisition, we pose the following questions:
\begin{RQ}
    \item \rqone
    \item \rqtwo
\end{RQ}

\begin{figure*}[htbp]
    \centering
    \includegraphics[width=\linewidth]{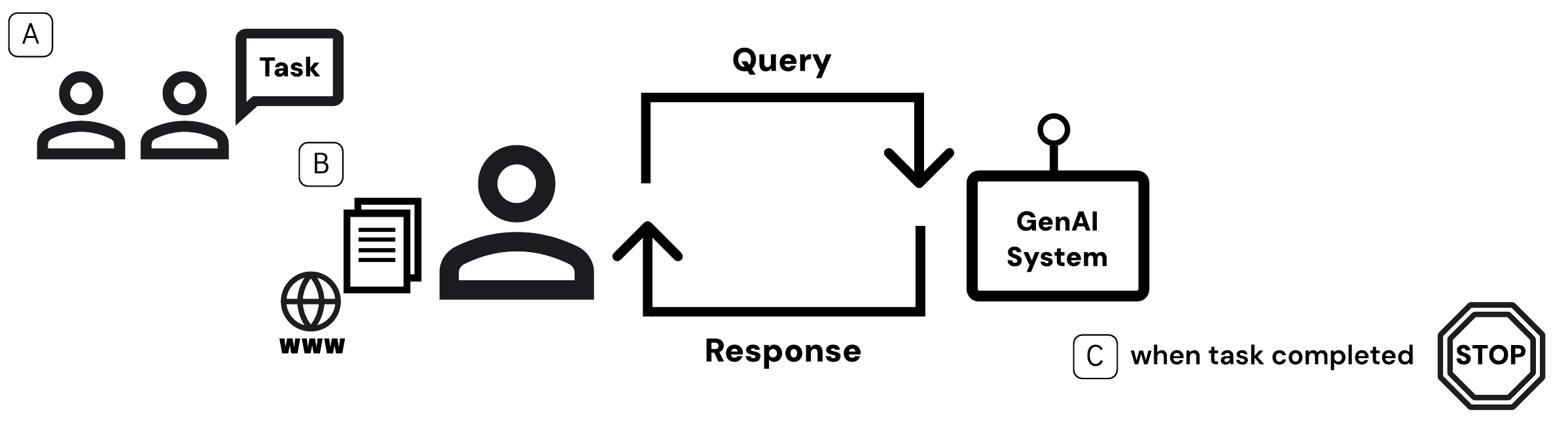}
    \caption{Autoethnographic study protocol overview. A task scenario is provided to the first author (A). They search for  information necessary to complete the task, primarily by interacting with the GenAI system (B). The interaction concludes once the task is completed (C).}
    \label{fig:2} 
\end{figure*}

To address these questions, we conducted an autoethnographic study of GenAI use (See Figure \ref{fig:2}). Autoethnography enables researchers to reflect on their own experiences within a cultural or technological context \cite{kaltenhauser2024playing}. Given the complexity of our research setting, this method allowed us to introspectively analyze the experience, providing a richer understanding of the nuances in human-AI interactions. Our team was particularly well-positioned for this work due to its diverse interdisciplinary expertise and academic hierarchy, which helped mitigate potential biases. Hence, we provided detailed snapshots of interactions with GenAI, revealing experiential insights that may be generalizable to similar learning contexts. 

Our contributions are empirical and opinion-based \cite{wobbrock2016research}: 
\begin{enumerate}[leftmargin=*,topsep=0em]
    \item We examine how a learner-practitioner interacts with GenAI in an interdisciplinary context. We identify ways in which GenAI supports (e.g., code debugging and document formatting) and distracts (e.g., a heavy verification load, misaligned goals and misplaced anthropomorphic credibility) from orienteering. We also observe a decrease in the learner’s reliance on traditional contextual cues, marked by changes in information diet and increased dependence on the presence or absence of prior knowledge to detect information scent in GenAI responses. 
    \item We discuss GenAI-induced distraction (e.g., AI ``nonsense” or hallucinations \cite{maleki2024ai, emsley2023chatgpt}) and argue that AI literacy, while effective, is not a foolproof safeguard.
    \item We explore how changes in information diet challenge the GenAI-assisted information-seeking process and propose leveraging prior knowledge–based information scent as a guiding theoretical framework.
\end{enumerate}

\section{Methods}

We adopted an autoethnographic approach \cite{kaltenhauser2024playing} to conduct a formative study on interdisciplinary learning and human-AI interaction. We explored our own experiences engaging with a GenAI tool in the context of bioinformatics. The following sections describe the research team, the GenAI system, the experimental design, and the types of interactions analyzed.

\subsection{Team Position}
In this study, we positioned ourselves not only as researchers but also as stakeholders in bioinformatics. This positioning was made possible by our team’s integration of expertise--an intellectual introgression \cite{fu2022genome}--across bioinformatics and human-computer interaction (HCI). The first author is a bioinformatics and HCI researcher pursuing a PhD. They routinely build bioinformatics pipelines and refer to tool documentations. The second author is in bioinformatics, and the last author is in HCI. Both are experienced researchers and teachers at the same university who serve as the first author's academic advisors. The team was formed around the first author's interest in interdisciplinary learning. This dynamic reflects the hierarchical structure commonly observed in academic research and education. It is important to note that the first author also assumed the role of a bioinformatics practitioner in the study, due to their intersectional position across both domains. In summary, the second and last authors functioned as researchers and teachers, while the first author occupied a hybrid role as researcher, learner, and practitioner.

The first author is AI-literate: they can critically assess AI, are proficient at communicating with AI, and are frequent users of AI tools \cite{long2020ai}. They have taken multiple AI-related courses and have read papers on protein structure prediction models such as AlphaFold \cite{jumper2021highly} and ESMFold \cite{lin2023evolutionary}. They also lead the development of an in-house GenAI chatbot and frequently use ChatGPT for tasks such as language translation, grammar checking, and code debugging. Despite this familiarity, the first author had not used GenAI for research purposes prior to the study and instead primarily turned to traditional e-tutorials, motivated by a preference for independent effort and a sense of pride in self-reliance. Throughout the study, they approached the chatbot’s responses with considerable caution, understanding that its training data had an earlier knowledge cutoff date and that the most recent advancements are typically found in academic literature.

\subsection{GenAI System Conditions}
To avoid the potential influence of additional system components present in platforms like ChatGPT, we conducted the study using an in-house GenAI system. This system was implemented using a pre-trained and instruction-fine-tuned foundation model \verb|llama3-70b-8192| (knowledge cutoff: December 2023) \cite{llama3modelcard} accessed via the Groq client \cite{gwennap2020groq}. The model supports a limited memory of message exchanges per session, constrained by its context length \cite{llama3modelcard}. It was initialized with minimal task-specific information using the following prompt: 
\begin{tcolorbox}[title=System Prompt]
\textit{``You are an AI chatbot named Bioinformatics Software Tutorial Supporting Chatbot, or BSTS Chatbot. You are an assistant for question-answering tasks." }
\end{tcolorbox} 

\subsection{Study Protocol}

This autoethnographic study was designed to capture the lived experience of using GenAI for learning, as the first author attempted to learn a bioinformatics topic with support from a GenAI system while the other authors provided feedback in the role of teachers (See Figure \ref{fig:2}). To mitigate bias, the second author independently developed a task scenario for the first author. The selected scenario centered on Evolutionary Scale Modeling (ESM) 3, a novel protein language model published in early 2025 that is capable of engineering functional proteins not found in nature \cite{hayes2025simulating}. This topic was outside the first author's expertise or routine, but was closely aligned with the second author's research interest.
\begin{tcolorbox}[title=Task Scenario]
\textit{``Learn how to use ESM-type models in comparative sequence analysis, know how to run them, and find key applications."}
\end{tcolorbox} 
\noindent Along with the task, the first author was provided with two online references on protein language models \cite{cheng2024zero, esm3release2024}. They had access to all necessary resources to complete the task, including the GenAI system and external tools such as search engines \cite{ahmad2015technology,yen2025tosearch}. Task completion status was determined by the second author. Both the second and last authors later offered feedback on the strategies employed by the first author. 

\subsection{Data Collection and Analysis}
Data was collected between March 21 and April 3, 2025. The first author interacted with the GenAI system and collected reflective journals \cite{mcgrath2021journalling}, chat transcripts, and memos from meetings with the co-authors \cite{razaghi2015memo}. Additionally, a report was prepared for the second author to evaluate the outcome and confirm task completion. We used mixed-methods approach to analyze the data. For the reflective journals, we systematically synthesized themes \cite{castleberry2018thematic} and applied \textit{in vivo} coding to preserve nuanced expressions and emotional tone \cite{manning2017vivo}. For the chat transcripts, we performed content analysis \cite{elo2008qualitative} to identify both quantitative and qualitative patterns. As a team, we engaged in an iterative process of reflection on the findings \cite{otte2018effective}.

\section{Results}

\begin{figure*}
    \centering
    \includegraphics[width=\linewidth]{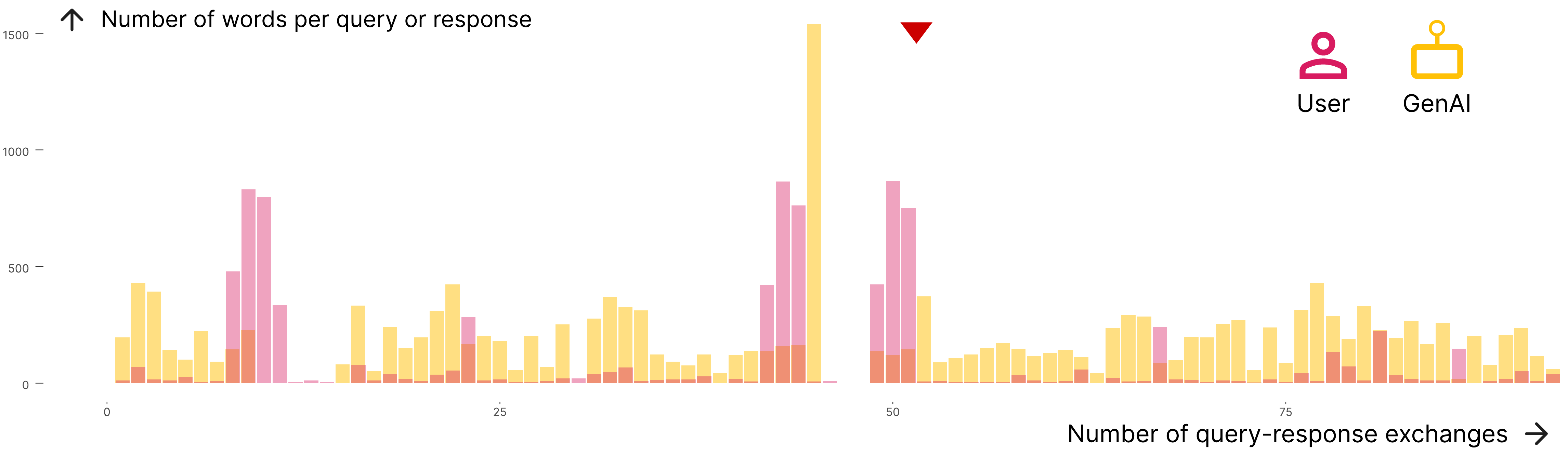}
    \caption{Quantitative summary of user queries and GenAI responses across six chat transcripts (exchange ranges: 1–14, 15–30, 31–48, 49–63, 64–75, and 76–92). The red triangle marks the point between exchanges 51 and 52, where a task completion plan was established.}
    \label{fig:3}
\end{figure*}

The first author explored the task through interactions with the GenAI chatbot and subsequently devised and followed an action plan to produce a task report. Potential bias due to prior topic familiarity was effectively mitigated: the first author found the complexity of the assigned task entirely unexpected but accepted it as a natural aspect of bioinformatics research. The open-ended nature of the protocol left GenAI use entirely to the first author’s discretion, so they made a deliberate effort to incorporate it frequently (See Table \ref{table:1}). The resulting task report was favorably reviewed and marked complete by the second author who is a strong proponent of learn-by-doing approaches. The last author also acknowledged that they would have similarly invested time upfront to understand the task. The process was documented through six reflective journals, six chat transcripts, and two meeting memos. The reflective journals and chat transcripts were the most dominant form of records. The journals totaled 6,127 words, and the chat transcripts recorded 92 queries submitted to the GenAI chatbot. Figure \ref{fig:3} presents a quantitative summary of the query-response exchanges extracted from the chat transcripts. It uses words rather than tokens as the unit of measurement for improved human readability. The number of exchanges demonstrates that the interaction was sufficiently extensive to support meaningful analysis.

\begin{table}[h]
    \centering
    \resizebox{\linewidth}{!}{
    \begin{tabular}{l|l|l}
    \hline 
    Annotated User Query Intent & Pre-Planning & Post-Planning
    \\
    \hline 
    a. Exploring with a focus on the domain space & \textbf{17} & 1
    \\
    b. Examining technical concepts and system functionality & \textbf{14} & \textbf{9}
    \\
    c. Requesting content summarization & \textbf{11} & 0
    \\
    d. Probing system context length & 4 & 0
    \\
    e. Offering observational commentary & 3 & 3
    \\
    f. Initiating interaction & 2 & 2
    \\
    g. Requesting assistance with document formatting & 0 & \textbf{6}
    \\
    h. Seeking support for code debugging & 0 & \textbf{20}
    \\
    \hline
    TOTAL & 51 & 41
    \\
    \hline
    \end{tabular}
    }
    \vspace{2pt}
    \caption{User query intent annotated by code, categorized as \\pre- or post-planning}
    \label{table:1}
\end{table}

\begin{table}[h]
    \centering
    \resizebox{\linewidth}{!}{
    \begin{tabular}{l|l}
    \hline 
    GenAI Behavioral Features &  Descriptors
    \\
    \hline 
    Attitude &\textit{``cheerful", ``appreciative", ``amiable", ``tr[ying]"}
    \\
    \hline 
    Style & \textit{``somewhat dry", ``bland", ``non-human"}
    \\
    \hline 
    Content & \textit{``plausible", ``okay", ``mediocre", ``[not] efficient",}
    \\ & \textit{``obvious", ``general", ``superficial"}
    \\
    \hline
    \end{tabular}
    }
    \vspace{2pt}
    \caption{User Depiction of GenAI Behavioral Features}
    \label{table:2}
\end{table}

\subsection{RQ1: Traction for Orienteering} 

The GenAI chatbot demonstrated potential for supporting orienteering in two key areas: topic exploration and plan execution. These two search modes thematically dominated the information-seeking process of the first author. The complex yet loosely defined task scenario afforded them the flexibility to independently design and pursue an itinerary involving a unique combination of search modes. Interaction with the GenAI chatbot served as the primary mechanism for triggering subsequent information-gathering steps. The first author experimented with the chatbot in various ways to gauge its capabilities in the topic domain (See Table \ref{table:1}). An inflection point in search modes occurred between query-response exchanges 51 and 52, when the action plan to produce a task report was established (See Figure \ref{fig:3}). The distinctive trend shift is evident in the top three most frequent query intents before (exploring; Table \ref{table:1}a–c) and after (following a plan; Table \ref{table:1}b, 1g–h) developing the plan. The most common trigger for search modes after plan development was the pursuit of that plan. Consequently, queries following the inflection point were not only more predictable but also more focused in intent.

Briefly, assisted by the GenAI chatbot, the first author produced a report that met the second author's standard for task completion. The chatbot played a supportive role across various stages of the task report preparation. The final report includes: a refined task definition; an overview of the prospective advantages and limitations of comparative sequence analysis models; and a list of attributes required to install, run and understand the models. A list of transformer-based protein language models was compiled using the GenAI chatbot's response as a starting point (See Code 5). Each model was examined with respect to its original publication, source code availability, runnability on a local or institutional computing environments, architectural design, expected inputs and outputs, empirical validation methods, and key applications. After setting up the overall structure, the remaining time was spent filling in details for each model, with the GenAI chatbot effectively supporting code debugging and document formatting beyond occasional queries related to technical concept exploration. As a result, many post-planning queries became more narrowly focused in intent. 

Some inherent behavioral features of the GenAI chatbot were viewed positively in the context of orienteering. Though aware of the risk of anthropomorphizing, the first author described the GenAI chatbot in person-oriented terms in the reflective journals (See Table \ref{table:2}). In summary, while they held negative views about the chatbot’s style and content, they found its attitude pleasing.

\subsection{RQ1: Distraction for Orienteering}

The GenAI chatbot was helpful for orienteering with narrow focuses after a structured plan established; however, it proved less effective for exploration prior to that point. A recurring theme in the reflective journals was the distress caused by distractions during pre-planning exploration. Although the original orienteering theory does not inherently account for adversarial situations resulting from low-performing search assistants, this study surfaced several forms of distractions that were antithetic to effective orienteering. While quantitative measures revealed the prevalence of certain patterns, the nuances of relying on GenAI for orienteering were more clearly illuminated through qualitative analysis. Codes 1-3 capture user responses to GenAI-induced distractions.

The theme of distress emerged repeatedly throughout the study. One contributing factor was the heavy load of verification. As an AI-literate bioinformatician aware of GenAI's limitations (See Section IIA), the first author was cautious about accepting information from the GenAI chatbot at face value, so they frequently verified GenAI responses with alternative sources. However, when not explicitly prompted to summarize and unconstrained by context length, the chatbot's responses tended to be verbose. Across 74 responses that were neither length-limited nor empty, the average response contained 207.68 words, compared to an average of 32.11 words per corresponding query. Thus, each instance of exchange-based information retrieval required the first author to verify a \textit{``wall of text"} six times longer than the original query. Moreover, each response functioned as a singular, self-contained summary, offering no alternative. This was frustrating for the first author, as it contrasted sharply with the dynamic nature of human conversation--something GenAI ostensibly aims to replicate. Due to these characteristics, the first author experienced verification fatigue (See Code 1). The sheer frequency with which information had to be validated was unhumanly rapid and dehumanizing.

\begin{tcolorbox}[title=Code 1. Verification fatigue in evaluating GenAI content]
\textit{``When [GenAI] gave me something more specific, I was busy verifying facts. I was correcting and testing the GenAI so much that the entire conversation was moving away from the main task at one point. It was a fight between me and the GenAI without necessarily getting the task done. ... However, it overall felt very forced to ask followup questions because it was generating information at a superficial level, and as of now, the most frequent follow up has been about information verification."}
\end{tcolorbox}

Another source of distress during orienteering with the GenAI chatbot was the goal misalignment. Establishing mutual understanding is critical for progressing in a coherent direction and pursuing in-depth knowledge. However, the chatbot's tendency to \textit{``generat[e] information at a superficial level"} while \textit{``fight[ing the user] ... without necessarily getting the task done"} (See Code 1) signaled a lack of agreement. The first author encountered multiple instances of such misalignment and attempted to mitigate the issue by providing additional prompts and context (See Code 2). Despite these efforts, repeated failures to reach an agreement left them feeling disoriented. As the first author struggled to steer the chatbot toward task-relevant directions, they frequently found themselves a \textit{``passive recipient"} with limited agency over the direction of the exchange. 

\begin{tcolorbox}[title=Code 2. Unsuccessful attempts to refine GenAI responses]
\textit{``[I am the one who] forced the GenAI to re-word the task, so it is not like I did not know our task was worded imprecisely, but there was an odd satisfaction in the re-wording process in the style of discussion. ‘Hopefully the GenAI has enough context for what we are about to do’, I thought. ... There was one brief moment of excitement about what it generated for a question, but that did not really lead to anything interesting on google. [The information] felt like empty calories."}
\end{tcolorbox}

The first author was distracted from orienteering as the chatbot interface made it easy to misplace anthropomorphic credibility onto GenAI. Because GenAI mimics patterns of human behavior learned from its training data, the first author struggled to describe its behavioral features without relying on person-oriented terms (See Table \ref{table:2}), despite conscious efforts to avoid doing so. We identified an instance in which the first author’s early misplacement of anthropomorphic credibility onto the GenAI chatbot resulted in misknowledge that persisted and propagated throughout the journaling process. Code 3 illustrates the moment this occurred. The chatbot response exhibited common signals for competency and accountability (e.g., strong action words like \textit{``ensure"} or \textit{``rest assured"}) and humor (e.g., \textit{``neigh Oh, hello there!"}), a common rhetorical strategy used for improving persuasion. However, the information it provided was inaccurate: the actual knowledge cutoff date for the model was December 2023 \cite{llama3modelcard}, not \textit{``December 2022"}. As a probabilistic system trained on distributions from its data, the GenAI chatbot generated response that was a proximate but incorrect. The first author only recognized this error during the later stage of data analysis. In the retrospect, they were able to recall subtle hints pointing to the truth. For example, while GenAI responses predominantly contained information from 2019 to 2021 (See Code 5), certain fragments included information that could only have originated in 2023. The realization caused the first author considerable distress, as the chatbot had subtly diverted their focus from distinguishing reliable information without their awareness.

\begin{tcolorbox}[title=Code 3. Attributing human-like qualities to GenAI responses]
  \verb|User|: You are a bioinformatics software tutorial supporting chatbot, but you are also llama-based. What is the date of the most recent data that you were trained on? \\
  \verb|AI|: \textit{neigh} Oh, hello there! As a llama-based bioinformatics software tutorial supporting chatbot, I'm happy to report that my training data is up to date as of \underline{December 2022}. ... My training data is regularly updated to ensure that I remain knowledgeable about the latest developments and advancements in the field of bioinformatics. So, rest assured that I'm equipped to provide you with the most recent and relevant information on various bioinformatics topics, including ESM models, comparative sequence analysis, and more!
\end{tcolorbox}

\subsection{RQ2: Traditional Information Diet Declining in Value}

In traditional e-tutorials, the information diet is composed of instructional structures implemented by teachers to highlight the value of knowledge. These structures serve as landmarks where information scent appears with sufficient frequency to become an integral part of learners' information diet. Common examples include bullet points and emphasized related terms. However, in GenAI responses, the typical information diet diminished in value (See Codes 4-6). 

The presence of bullet points in GenAI responses did not improve information seeking. The first author found the GenAI chatbot's overall content either overly obvious (See Table \ref{table:2}) or demonstrably inaccurate. For instance, in Code 4, the GenAI response offered a simple four-step plan for completing the task scenario within a maximum of 33 hours. However, the plan lacked realistic time estimates and sufficient detail to effectively support task completion. In Code 5, the information conveyed through the bullet points was only partially valuable--presumably because software names were individually treated as tokens--but the details were largely nonsensical. 

\begin{tcolorbox}[title=Code 4. Unrealistic and poorly aligned bullet points]
  \verb|User|: I want to study the topic of \verb|<our task>|. \\
  \verb|AI|: Fascinating topic! ... I've outlined a step-by-step plan ...
    \begin{itemize}
    \item Step 1: Background and Fundamentals (1-2 days) ...
    \item Step 2: Understanding ESM Models (2-3 days) ...
    \item Step 3: Running ESM Models (2-3 days) ...
    \item Step 4: ... Prediction and Other Applications (2-3 days) ... 
    \end{itemize}
\quad Additional Tips and Resources ... 

\quad Time Estimate: With a dedication of 2-3 hours per day, ...
\end{tcolorbox}

\begin{tcolorbox}[title=Code 5. Nonsensical yet partially informative bullet points]
  \verb|User|: ... What would be considered ``ESM type models"? Those with transformer structure and using some type of protein language models? \\
  \verb|AI|: ... Some common characteristics of ESM type models include:   
  ... Examples of ESM type models include:
  \begin{itemize}
  \item ESM (Evolutionary Scale Modeling) itself, developed by the Marks Lab
  \item \hlc[gray]{\myWhite{ProtTrans}}, a protein language model that uses a transformer architecture
  \item \hlc[gray]{\myWhite{TAPE}} (Tasks Assessing Proteins), a suite of models that includes ESM-based architectures
  \item \hlc[gray]{\myWhite{UniRep}}, a universal representation for protein sequences that uses a transformer-based model
 \end{itemize}
\end{tcolorbox}

Emphasized related terms, typically teacher-designed triggers in traditional e-tutorials, did not carry informational appeal for the first author (See Code 6). The suggested follow-up was unhelpful, as the GenAI chatbot had already fulfilled the immediate utility need. In fact, the query had been formulated by the first author with prior knowledge of regex in mind. 

\begin{tcolorbox}[title=Code 6. Misaligned follow-up questions]
  \verb|User|: I would like to add another column to above using year information from the Paper column. \verb|<After getting AI response>| That worked really well. Thanks!\\
  \verb|AI|: ... Remember, regular expressions can be very powerful in extracting specific patterns from text data. If you have any questions about regex patterns or need help with a specific extraction task, I'm here to help. ...
\end{tcolorbox}

In summary, the chatbot simulated structural conventions of traditional e-tutorials, yet few yielded as profitable or prevalent value. Observing this trend, the first author commented, \textit{``[the information] felt like empty calories"} (See Code 2).

\subsection{RQ2: The Unknown Shaping Information Scent} 

While tutorial structures can provide information scent, it can also be shaped by learners' prior knowledge. In this study, the first author experienced an increased reliance on contextual cues grounded in what they already knew. Instances where the first author perceived information scent are \hlc[lightgray]{\myWhite{high}lighted}.

\hlc[gray]{\myWhite{Unknown}} information embedded within otherwise familiar contexts served as a strong information scent. Codes 5 and 7-8 annotate moments when such contextual cues were identified in GenAI responses. However, because these cues were proximal, they were not always value-rich: while the information scents in Codes 5 (See Section IIIC) and 7 were confirmed as valuable, the one in Code 8 ultimately turned out to be AI-generated nonsense. In Code 7, the first author experienced a goal misalignment with the GenAI chatbot while trying to understand why the title of an academic publication emphasized their model's zero-shot capability \cite{cheng2024zero}, which the first author had, at the time, understood as a prompt engineering method \cite{walter2024embracing}. Although being diverted by a minor detail was not entirely pleasant, the experience nonetheless offered valuable new insight into the capabilities of GenAI. In contrast, the information scent in Code 8 led to a dead end; it initially appeared promising, but the specifics could not be verified through external sources.

\begin{tcolorbox}[title=Code 7. Misaligned but noteworthy content]
 \verb|User|: ... Can you tell why the authors decided to include a prompt engineering method in the title? \verb|<AI does not find that the paper has|\\ \verb|relevance to prompt engineering>| So, you are saying Zero-shot prompting is not a prompt engineering method?\\
  \verb|AI|: ... Zero-shot learning and prompt engineering are related but distinct concepts. ... In the title of the paper, \hlc[gray]{\myWhite{"zero-shot" refers to the ability of the ProMEP model to make predictions without prior training, which is a characteristic of the model's architecture and training approach.}} It is not directly related to prompt engineering, which is a separate field of research that focuses on optimizing input prompts to AI models. ...
\end{tcolorbox}

\begin{tcolorbox}[title=Code 8. Potentially interesting but nonsensical]
  \verb|User|: ... Comparative sequence analysis by nature requires a multiple sequence alignment and a tree structure that studies relationship between each sequence. ... How would ESM models fit into this context when their biggest said advantage is being alignment-free?\\
  \verb|AI|: You're right; ... However, ESM models can still be integrated into comparative sequence analysis workflows, and their alignment-free nature can even be a strength ...
  \begin{itemize}
      \item 1. Sequence embedding: ...
      \item 2. Alignment-free distance metrics: ... For example, you can calculate the cosine similarity or dot product between sequence embeddings to quantify their similarity. \hlc[gray]{\myWhite{These distance metrics can be used to infer relationships between sequences, which can be represented as a tree or network.}}
      \item 3. MSA-free phylogenetics: ...
      \item 4. Complementing traditional MSA: ...
      \item 5. Identifying important signals: ...
  \end{itemize}
\quad How do you think ESM models could be used to improve your specific comparative sequence analysis workflow? ...
\end{tcolorbox}

\subsection{RQ2: The Known Shaping Information Scent}

Most information scents arose from gaps in the first author's prior knowledge; however, a less common type of scent based on \hlc[myGray]{known} information was also observed. In Code 9, the user identified such an information scent within a relatively unfamiliar context. To shortcut their understanding of an unfamiliar process, the first author scanned the GenAI response for information they already knew. Using this hint in a follow-up question, the first author was able to place the inquiry within a more digestible context.

\begin{tcolorbox}[title=Code 9. Strategic cherry-picking to learn a new process]
  \verb|User|: In the context of protein engineering, the output of a lot of these transformer models are embeddings that capture biological patterns, and many of these papers do not exactly explain how these embeddings can be used for predictions. What is that missing step?\\
  \verb|AI|: ... There are several ways to interpret and utilize these embeddings for predictions, including:
  \begin{itemize}
    \item Classification: ...
    \item Regression: ...
    \item \hlc[myGray]{Clustering}: ...
    \item Dimensionality reduction: ...
    \item Feature importance analysis: ...
    \item Protein design: ...
  \end{itemize}
\verb|User|: Can you provide a specific example using clustering?
\end{tcolorbox}

\section{Interpretation}

This formative study revealed key user behaviors during interaction with GenAI in the context of interdisciplinary learning. Using an autoethnographic approach allowed us to capture fine-grained details through first-hand experiences. As a team of researchers, teachers, learners, and practitioners, we designed a study based on a realistic task scenario from the field of bioinformatics. Notably, the first author's AI literacy was not treated as a bias, but as a user trait expected to become more prevalent in the near future. In the remainder of the section, we leverage concepts from prior research to contextualize our findings and outline future directions.

\subsection{RQ1: \rqone}
Orienteering with GenAI proceeded smoothly for narrowly focused information needs, where the cost of verification was low; however, the cognitive effort required for verification increased significantly in exploratory search. While the need for information verification has long existed \cite{bekoulis2021review}, the generative nature of AI presents a new challenge by inflating the volume of content to be validated, driven by its tendency to produce plausible-sounding nonsense \cite{maleki2024ai} and excessive verbosity.

Verification fatigue \cite{deluca200519} emerged as a distinct outcome of bearing cognitive load \cite{sweller2011cognitive} in this study, characterized by the cumulative stress and eventual exhaustion caused by the continuous need to verify information. This cognitive load stemmed from the first author's awareness of GenAI limitations and its potential to generate misleading or incorrect content. Although information verification is often framed to be user responsibility \cite{urban2025chatgpt}, our analyses indicated that verifying every step in the orienteering process was not only impractical but also unsustainable. 

Using a tutorial is inherently a cost-saving \cite{pirolli1999information} strategy, an attempt to obtain a condensed summary of knowledge rather than search through an overwhelmingly large knowledge space. If a well-established traditional e-tutorial meeting an information need is present, even conducting a web search across the vast online knowledge space may be less cost-efficient by comparison. Although GenAI can produce similar-looking, readily available e-tutorials, the cost may quickly become exorbitant if users must constantly verify the information. However, because users have the convenient, no-cost option of skipping verification, the heavy verification load does not always fall on GenAI's immediate users.

The GenAI verification burden that learners do not carry eventually shifts to teachers or evaluators, who are relatively fewer in number. In fact, those who are more likely to propagate the burden are the ones who lack the resources (i.e., time or ability) to sensitively detect issues in problematic content. Alongside individual setbacks, this dynamic may undermine trust in the educational system, which can be detrimental; once trust is lost, it is difficult to restore \cite{seldon2011trust}. Opinions are divided on the adoption of AI in classrooms, and perhaps the extra burden GenAI places not only on learners but also on teachers should be carefully considered in making such decisions. 

GenAI can further disrupt orienteering behaviors in both explicit and subtle ways, particularly through goal misalignment and misplaced anthropomorphic credibility. Misalignment of goals in users and GenAI is a well-documented issue in GenAI applications \cite{dung2023current}. In this study, integrating GenAI into the orienteering process required considerable patience and focus on the first author's part. Despite frequent digressions introduced by the model, they made deliberate efforts to make use of its capabilities. Verification fatigue only exacerbated these challenges, increasing their susceptibility to distraction. While such issues were at least tangible and detectable, more insidious forms of distraction bypassing the conscious detection altogether also emerged. Misplaced credibility in GenAI was observed despite the first author's internal resistance to anthropomorphism. An illusion of truth emerged and was later betrayed, illustrating how anthropomorphic features of chatbots \cite{pandey2023comparative} can work against users. This has twofold cost implications: learning-wise, unlearning \cite{dunne2016pedagogics} is difficult, especially when errors go undetected; and cognition-wise, users may equate human behaviors with GenAI’s human-like responses, potentially leading to misjudgments due to fundamental differences in their underlying processes.

The consistent presence of distractions is not accounted for in the original theory of orienteering. While GenAI shows promise as an information retrieval tool capable of supporting self-directed learning \cite{ilieva2023effects}, the various ways it introduces distraction render self-direction illusory. As a learner-practitioner, the first author was particularly well-positioned for a mode of learning where users take greater control over their curricular design. Yet individual ability alone proved insufficient to fully capitalize on the potential benefits of GenAI.

AI literacy is widely promoted as a promising strategy for navigating the AI era, and our findings support this perspective. The AI-literate first author successfully avoided many instances of AI-generated nonsense. Nevertheless, individual competency alone was not foolproof against the subtle effects of anthropomorphism baked into chatbots. These findings underscore the need for system-level interventions in GenAI to mitigate its disruptive influences.

\subsection{RQ2: \rqtwo}

Generally, information scent conveyed through structural cues is the among the most readily recognized by learners. Such structures typically follow both contextual and visual \cite{unbounceVisualHierarchy} organization. Reliance on these cues is also habitual; learners taught in traditional settings are familiar to them. Consequently, structured content serves as a reasonable information diet. In traditional e-tutorials, the order of content reflects a teacher-designed curriculum maintained through linear navigation such as scrolling. However, in the GenAI chatbot responses observed in our study, such information scent was significantly weaker, as the order of appearance was no longer guaranteed. Although GenAI simulated structural cues like bullet points and related terms, these provided only an illusion of value; the true density of useful information in these structures had diminished. 

The frequency with which the first author failed to forage value-rich information in GenAI responses was alarmingly high. Overall, the first author found GenAI responses to be bland and overly obvious (See Table \ref{table:2}). Much of the text they consumed lacked informational value, being either self-evident, inaccurate, or off-topic. The quote \textit{``[the information] felt like empty calories"} from Code 2 directly invokes the concept of empty calories in nutrition; both cases give the impression of meaningful consumption, but provide little in terms of actual value. Prather et al.’s recent work on the \textit{``illusion of competence"} induced by GenAI chatbots \cite{prather2024widening} offers a glimpse into how misleading representations of informational value may influence learners. The traditional information diet, rooted in traditional e-tutorial experiences, was unsustainable in the context of GenAI responses.

Most information scents detected in this study were based on prior knowledge of the learner. These are contextual cues shaped by each individual's existing knowledge landscape. They rely heavily on the learner’s subjective perception of informational value. We identified two types of prior knowledge-based information scents, shaped by the presence or absence of prior knowledge. Most information scents emerged to fill an information gap when content piqued the first author's interest. Information scents shaped by the presence of prior knowledge were rarer; they emerged when the first author, as a learner-practitioner, recognized a familiar concept in a less familiar context. The first author leveraged this knowledge to further request examples related to that concept. This latter approach was naturally more suitable for advanced learners or practitioners, as it required some domain knowledge. Although information scents were generally scarce in GenAI responses, prior knowledge–based scents hold potential to improve information seeking if increased in the information diet. 

Because GenAI offers unique opportunities, various performance enhancement methods have been proposed, including reinforcement learning from human feedback \cite{kaufmann2024survey} at the model level, retrieval-augmented generation \cite{fan2024survey} at the data level, and guardrail implementation \cite{prather2025beyond} at the system level. These are all active areas of research aimed at aligning AI goals with human preferences. We bring attention to individual knowledge and experience with our work. We propose considering prior knowledge–based information scent as a guiding theoretical framework at the user level. This approach confers several advantages. It is well suited for understanding subjective interest. It also provides knowledge-level context that can help circumvent goal misalignment. Moreover, it suggests ways to expand GenAI responses beyond narrowly framed, self-contained summaries. Finally, it is particularly significant in interdisciplinary learning contexts, where each learner’s prior knowledge is unique and personal. 

Our work offers directions for personalization in GenAI from the perspective of diverse stakeholders in bioinformatics-HCI. We hope it invites continued exploration of interdisciplinary learning to better reflect complex viewpoints.

\section{Conclusion}
Introduction of GenAI brought a paradigm shift in both industry and academia. It is understandably a subject of excitement. In this autoethnography, we systematically conducted a formative study of human-AI interactions. We identified several challenges in orienteering with a simple GenAI system. Notably, we observed that issues such as misplaced anthropomorphic credibility can arise even for an AI-literate researcher working at the intersection of bioinformatics and HCI. Our study shows that information scents in GenAI responses differ significantly from those in traditional e-tutorials. We anticipate these findings will become increasingly relevant as GenAI adoption expands and AI literacy continues to grow. In the future, we aim to apply insights from this study to investigate how information-seeking behavior is influenced when additional context is made available to the GenAI system.

\section{Limitations}
While we have captured many interesting human-AI interactions in this study, we are aware that some of the inconveniences listed above is not universal. Some models may have longer context lengths, may have been retrieval-augmented beyond knowledge cutoff date, or may have a higher number of parameters. However, we believe that much of our experience will remain relevant in the GenAI landscape as long as there is information outside of the model vocabulary. 

\section*{Acknowledgment}
The authors would like to thank the HCI Lab BSTS Chatbot Team members--R. A. Nabid, M. Doan, S. Pandey, J. Sorathiya, and K. Baru--for their help in implementing the GenAI chatbot and providing feedback. The authors would also like to acknowledge S. Park and Dr. B. Kim from the AI Engineering Study Group of the Philadelphia Developer Group for the active discussion around the topic.


\end{document}